\documentclass[11pt,a4paper]{article}

\usepackage{epsfig}
\usepackage{latexsym}

\newcommand{\bi}{\begin{itemize}}
\newcommand{\ei}{\end{itemize}}
\newcommand{\be}{\begin{equation}}
\newcommand{\ee}{\end{equation}}

\newcommand{\bea}{\begin{eqnarray}}
\newcommand{\eea}{\end{eqnarray}}
\newcommand{\beastar}{\begin{eqnarray*}}
\newcommand{\eeastar}{\end{eqnarray*}}

\newcommand{\eq}[1]{~(\ref{#1})}
\newcommand{\eqq}[2]{~(\ref{#1},\ref{#2})}

\begin{document}

\title{Universal dependence of the fluctuation-dissipation ratio on the transition rates
in trap models}

\author{F Ritort~${\footnote{On leave at Department of Physics, University of California, Berkeley CA 94720,
USA}}$\\
Department of Physics, Faculty of Physics,
University of Barcelona\\ Diagonal 647, 08028 Barcelona, Spain\\
{\tt ritort@ffn.ub.es}}

\maketitle

\abstract{We investigate violations of the fluctuation-dissipation
theorem in two classes of trap models by studying the influence of the
perturbing field on the transition rates. We show that for perturbed
rates depending upon the value of the observable at the arrival trap, a
limiting value of the fluctuation-dissipation ratio does exist.
However, the mechanism behind the emergence of this value is different
in both classes of models. In particular, for an entropically governed
dynamics (where the perturbing field shifts the relative population of
traps according to the value of the observable) perturbed rates are
argued to take a form that guarantees the existence of a limiting value for the effective
temperature, utterly related to the exponential character of the
distribution of trap energies.  Fluctuation-dissipation (FD) plots
reproduce some of the patterns found in a broad class of glassy systems,
reinforcing the idea that structural glasses self-generate a dynamical
measure that is captured by phenomenological trap models.}

\section{Introduction}

Glasses are widespread materials that yet resist a basic comprehension
from many points of view. Although several aspects concerning their
macroscopic behavior have been modeled, showing qualitative agreement
with experimental results, still our understanding of the relevant
physics at the mesoscopic scale remains largely speculative. Long
timescales (i.e. several orders of magnitude larger than the microscopic
timescales) associated to physical processes occurring at very small
length scales (of the order of several nanometers) are thought to be
common. The main difficulty behind a theoretical description of glassy
dynamics is caused by the great disparity between timescales and length
scales, where neither hydrodynamic approaches or standard nucleation
theories can be applied.  It is for this reason that trying to
understand the dynamical behavior of glasses remains a theoretical
challenge. Its comprehension may open new routes to the understanding of
a large class of new physical phenomena, where such a strong disparity
of temporal and spatial scales is relevant.

The glass transition problem has been studied from many perspectives. In
the absence of a well established basic microscopic theory modeling
offers many possibilities.  A successful category of models for the
glass transition are trap models~\cite{BM,BMM}. They have the advantage
of capturing some of the relevant features of glassy relaxation within a
tractable formalism. Trap models have been widely investigated (for a
review, see \cite{BouCugKurMez98}) and found to reproduce pretty well
several experimental features such as the decay of the thermoremanent
magnetization in spin glasses~\cite{BouDea95,BerBou02} or the
super-activated behavior of the viscosity in glasses~\cite{BouComMon95}.
Glassy systems are characterized by aging and violations of the
fluctuation-dissipation theorem (FDT) (for recent reviews see
\cite{CriRit03,Cugliandolo02}). FDT violations can be quantified by
defining the fluctuation-dissipation ratio (FDR),
\be
X(t,s)=\frac{TR(t,s)}{\frac{\partial C(t,s)}{\partial s}}~,~~~~t>s
\label{a1}
\ee
where $T$ is the temperature of the bath (the Boltzmann constant $k_B$
is set to 1) and $C(t,s), R(t,s)$ are the two-times correlation and
response functions. The FDR allows to define also the effective
temperature,
\be
T_{\rm eff}(t,s)=T/X(t,s)~~~~.
\label{a1b}
\ee
In equilibrium FDT is satisfied so $X=1$ and $T_{\rm eff}(t,s)=T$. These
definitions would be devoid of any interest if it were not for the fact
that a broad category of glassy systems (those exhibiting a one-step
replica-symmetry breaking transition in their corresponding mean-field
theory) show, in their long-time regime, that $X(t,s)\to \hat{X}(s)$,
i.e. the FDR converges to a non-trivial function of the lowest time $s$.
More precisely, these systems show the existence of two-timescales
depending upon the value of the ratio $(t-s)/s$ in the asymptotic regime
$t,s>>1$. If $(t-s)/s<<1$ then $X(t,s)=1$ and the systems is locally
equilibrated (short-time sector). However, if $(t-s)/s \sim {\cal O}(1)$
then $X(t,s)\to \hat{X}(s)$ (long-time sector) being independent on the
precise value of the ratio $t/s$ (provided it is finite). Hereafter, we
will refer to this category of models as one-step (1S) models as they
show a single jump of the value of $X(t,s)$ from 1 to a different value
(generally smaller than 1). The existence of $\hat{X}(s)$ shows that the
aging state at time $s$ can be characterized by a non-equilibrium
parameter that could be identified as the fingerprint of a
non-equilibrium temperature, $T_{\rm eff}(s)=T/\hat{X}(s)$.  From a
theoretical point of view, the conditions necessary for the existence of
$\hat{X}(s)$ in the long-time sector have not been found in full
generality. These conditions could be quite strong, though, and the
possibility has been raised that only the limiting value
$X_{\infty}(s)=\lim_{t\to\infty}X(t,s)$ (rather than $\hat{X}(s)$) is
physically meaningful~\footnote{The statement has been done in the
context of ferromagnetic models at criticality~\cite{GodLuc00a} where
the universal FDR is
$X_{\infty}=\lim_{s\to\infty}X_{\infty}(s)$. Several analytical works
support the conjecture~\cite{LipZan00,CalGam02}.  The conjecture can be
naturally extended beyond criticality to generic glassy systems by
proposing that $X_{\infty}(s)$ is waiting-time dependent.}.  Most of the
numerical studies done on structural glass models seem to be in
qualitative agreement with the 1S scenario \cite{CriRit03}.

Usually, FD violations are schematically depicted in the form of FD
plots where the integrated-response function
$\chi(t,s)=\int_s^tR(t,t')dt'$ is plotted as a function of 
$C(t,s)$. 
For 1S models FD plots are made out of two linear segments, one of slope
 $-1/T$ covering the short-time sector, the other with slope
$-\hat{X}(s)/T$ in the long-time sector.

Recently, a debate has emerged whether trap models can reproduce the
pattern characteristic of FD violations in 1S systems. In particular,
because dynamics within traps is usually not incorporated into the
models, they should only reproduce the long-time sector where
$\hat{X}(s)$ describes relaxational jumps among traps.  Yet, the most
recent results on this issue~\cite{FieSol02} seem to
disprove the existence of such 1S features.

The goal of this paper is to clarify this issue by showing that FD plots
in trap models strongly depend upon the functional dependence of the
transition rates on the field. Actually, perfectly straight 1S FD plots
can be obtained provided the escape rate from a trap is unchanged
(within the linear-response regime) by the action of a external
perturbation. The study of two different classes of trap models shows
the existence of a value $X_{\infty}(s)$ for the FDR \eq{a1} in the
limit $t\to\infty$, yet the mechanism behind the emergence of this
limiting value is different.  For a dynamics governed by entropic
effects there exists an effective temperature $T_{\infty}(s)$ that
remains finite in the limit $T\to 0$ and is uniquely determined by the
width of the exponential distribution of trap energies. This shows that
1S behavior with $X_{\infty}(s)$ proportional to $T$ at low $T$ is the
fingerprint of a slow dynamics governed by entropic rather
than thermally activated effects.

\section{Trap models: some definitions and useful expressions}
\label{definitions}

Following \cite{BM} we consider $N$ independent traps labeled by the
letter $\alpha$, each trap characterized by an energy $E_{\alpha}$ and a
generic observable that we will denote by $M_{\alpha}$. We consider a
Markov dynamics among traps. To set the formalism as clear as possible
we will consider time as discrete and define a dynamical history by the
temporal series $\lbrace\alpha(t)\rbrace$ where $t$ is an integer
(and consecutive time steps are separated by $\Delta t=1$). The
dynamical evolution is specified by the probability $P_t(\alpha)$ that
the system stays in trap $\alpha$ at time $t$, and by the conditional
probability $W(\alpha'|\alpha)$ that the system goes from trap $\alpha$
to $\alpha'$ in a single time step. We assume conditional probabilities
to be time-independent. Eventually we will be interested in the
continuous-time limit $\Delta t \to 0$ where conditional probabilities
for $\alpha\ne \alpha'$ are vanishingly small $W(\alpha'|\alpha)\to
\hat{W}(\alpha'|\alpha)\Delta t$, $\hat{W}(\alpha'|\alpha)$ being the
transition rates. In what follows we will coin the term rate for $W$
although properly speaking this term should be only adopted for its
continuous-time analogue $\hat{W}$.  A master equation governs the
dynamical evolution of the system,
\be
P_t(\alpha)=\sum_{\alpha'}P_{t-1}(\alpha')W(\alpha|\alpha')~~~.
\label{a2}
\ee
 The
$W$ will consist of two terms, the term $W(\alpha'|\alpha)=A(\alpha'|\alpha)$ for
$\alpha\ne \alpha'$ describing the jumping rate from $\alpha$ to
$\alpha'$ and the term $W(\alpha|\alpha)=B(\alpha)$ describing
the rate of staying in trap $\alpha$. Therefore we can write,
\be
W(\alpha'|\alpha)=A(\alpha'|\alpha)(1-\delta_{\alpha,\alpha'})+B(\alpha)\delta_{\alpha,\alpha'}
\label{a3}
\ee
Transition rates are normalized $\sum_{\alpha}W(\alpha|\alpha')=1$ and satisfy,
\be
\sum_{\alpha'\ne \alpha}A(\alpha'|\alpha)+B(\alpha)=1, ~~~\forall \alpha~~~~.
\label{a4}
\ee
The transition rates allow to define the escape rate $W_{\rm
esc}(\alpha)$ from a given trap $\alpha$,
\be
W^{\rm esc}(\alpha)=\sum_{\alpha'\ne \alpha}A(\alpha'|\alpha)= 1-B(\alpha)
\label{a5}
\ee
Dynamics in trap models is specified once the rates $W$ are given. For
the time being we concentrate only in the zero-field case where dynamics
is only governed by the energies of the traps.  We consider two
representative cases: the Bouchaud model (BM) \cite{BM} where the rate
$A(\alpha'|\alpha)$ only depends on the energy $E_{\alpha}$ of the
departure trap $\alpha$, and the Barrat-Mezard model (BMM) \cite{BMM}
where the rate $A(\alpha'|\alpha)$ depends on both energies of the
departure and arrival traps, $E_{\alpha}$ and $E_{\alpha'}$. Both models
consider an spectrum of traps with energies $E_{\alpha}\le 0$ and are
defined by their respective rates,
\bea
A^{\rm BM}(\alpha'|\alpha)=\frac{\omega_0}{N}\exp(\beta E_{\alpha})\label{a6a}\\
A^{\rm
BMM}(\alpha'|\alpha)=\frac{\omega_0}{N}\frac{1}{1+\exp(\beta(E_{\alpha'}-E_{\alpha}))}\label{a6b}
\eea
where $\beta=1/T$ and $\omega_0$
is a microscopic frequency that we will take inversely proportional to
the discretization time $\Delta t$.  The choices \eqq{a6a}{a6b} are justified by the
requirement that the detailed balance property is satisfied. The
corresponding expressions for $B(\alpha)$ are determined by the
normalization condition \eq{a4}. The energy spectrum of the ensemble of traps
is defined by,
\be
\rho(E)=\frac{1}{N}\sum_{\alpha=1}^N \delta(E-E_{\alpha})
\label{a7}
\ee
Both models consider an spectrum of states with exponentially
distributed energies, 
\be
\rho(E)=\frac{1}{T_g}\exp\Bigl(\frac{E}{T_g}\Bigr)~~~~~E\le 0~~~~~.
\label{a7b}
\ee
There is an important difference between the dynamics in the BM and the
BMM in the low $T$ limit. While dynamics is completely arrested in the
BM (the escape rate \eq{a5} vanishes exponentially fast with $1/T$) it
does not in the BMM where the escape rate is finite even at
$T=0$~\footnote{This difference between both models appears explicitly
in the time decay of the energy in the glassy phase $T<T_g$ where the
Gibbs energy distribution ceases to be normalizable. While in the BM the
energy decays as $E(t)\sim -T\log(t)$ (therefore relaxation arrests at
$T=0$), in the BBM it decays like $E(t)\sim -T_g\log(t)$ for $T\to 0$
being $T$ independent in that limit.}. This makes the BM model more
suitable to describe glassy relaxation in a landscape of traps separated
by energy barriers while the BMM is better suited to describe relaxation
over entropy barriers.  

In the presence of an external field $h$ coupled to the observable
$M_{\alpha}$ the rates \eqq{a6a}{a6b} get modified. The simplest
possibility is to assume that the functional form of the rates
\eqq{a6a}{a6b} remains unchanged while replacing 
\be
E_{\alpha}\to
E_{\alpha}-hM_{\alpha}.
\label{a7c}
\ee
We will refer to this choice as the {\em purely
activated} rates. A priori, however, other choices maybe equally valid. For
instance, let us consider the following rates in the presence of a
field~\footnote{From now we will use the subscript $h$ to refer to
field-dependent rates. For zero-field rates we will drop off the
subscript $h$.},
\be
A_h(\alpha'|\alpha)=A(\alpha'|\alpha)\exp\Bigl(-\beta h(\mu M_{\alpha}-\gamma M_{\alpha'})\Bigr)\label{a8}
\ee
where $\mu$ and $\gamma$ are arbitrary parameters. In this form these
rates generally violate detailed balance. Although this choice may seem
unjustified, violation of detailed balance may be a necessary ingredient
for trap models to reproduce non-equilibrium aspects of glassy systems~\footnote{\label{foot1}
Actually, trap models can be justified only as a coarse-grained
description where the original phase space is partitioned into
components or regions, each of these regions corresponding to a trap
(for a discussion of this aspect see Secs.~3.3 and 4.2 in
\cite{CriRit03}). However, at difference with the original microscopic
dynamics, the projected dynamics among traps can violate detailed
balance (a property that must be satisfied by the microscopic dynamics
for the system to equilibrate).}.

The rates\eq{a8} were considered in~\cite{BouDea95} in the case where
$\gamma=1-\mu$ and detailed balance is preserved. However, the
implications and physical meaning of the term $\gamma M_{\alpha'}$ in
\eq{a8} went unnoticed.  Indeed, the choice of purely activated rates
($\mu=1,\gamma=0$) has occupied the majority of studies of FDT
violations in trap models \cite{BouDea95,FieSol02,BerBou02}.  Here we go
beyond and show that the existence of the term $\gamma M_{\alpha'}$ is
crucial for the emergence of 1S behavior in FD plots.  We will give also
a physical interpretation of this term in a relaxational scenario
governed by activated and entropic barriers. Our study will consider both the BM and
the BMM.

Before delving into each of the two class of models we present some
useful formulae needed to derive the FDR.  Some of these computations
have been addressed in previous works~\cite{BouDea95,FieSol02,BerBou02},
here we reproduce them in more detail. We are interested in computing
the response function $R(t,s)$ and express it in terms of derivatives of
the correlation function $C(t,s)$. They are defined by,
\be
C(t,s)=<M(t)M(s)>~~~~;~~~~R(t,s)=\frac{1}{\Delta t}\frac{\delta <M(t)>}{\delta h(s)}\theta(t-s)~~~,
\label{a9}
\ee
i.e. the response function is the change of the expectation value
$<M(t)>$ when a impulse field coupled to $M$ is applied at a previous time $s$.
For simplicity we will assume that $C(t,t)$ is normalized to unity.
If $W_h(\alpha'|\alpha)$ stands for the rate in the presence of a field
then we have,
\be
R(t,s)=\frac{1}{\Delta t}\sum_{\alpha_t,..,\alpha_s}P_s(\alpha_s)M_{\alpha_t}\prod_{r=s+1}^{t-1}
W(\alpha_{r+1}|\alpha_r) \Delta_h W(\alpha_{s+1}|\alpha_s)
\label{a10}
\ee
where
\be
\Delta_h W(\alpha'|\alpha)=\lim_{h\to 0} \frac{W_h(\alpha'|\alpha)-W(\alpha'|\alpha)}{h}
\label{a11}
\ee
Using \eqq{a3}{a4} we can rewrite \eq{a11} as follows,
\be
\Delta_h W(\alpha'|\alpha)=a(\alpha'|\alpha)-\Bigl[\sum_{\alpha''}a(\alpha''|\alpha)\Bigr]\delta_{\alpha',\alpha}
\label{a12}
\ee
with the definition,
\be
a(\alpha'|\alpha)=\Bigl( \frac{\partial A_h(\alpha'|\alpha)}{\partial h}\Bigr)_{h=0}~~~.
\label{a13}
\ee
The next sections describe how to compute $a(\alpha'|\alpha)$ in both
the BM and BMM and how an expression for $R(t,s)$ in \eq{a10} can be obtained.


Another important concept related to the measured observables
$M_{\alpha}$ is neutrality~\cite{FieSol02,CriRit03}. The probability density of being at time
$t$ in a trap with a value $M$ of the observable, conditioned by its
energy to be equal to $E$, is given by,
\be
P_t(M|E)=\frac{\sum_{\alpha}P_t(\alpha)\delta(M-M_{\alpha})\delta(E-E_{\alpha})}{\sum_{\alpha}P_t(\alpha)\delta(E-E_{\alpha})}~~~.
\label{a14}
\ee
$M$ is said to be neutral if, assuming that
$P_{t=0}(M|E)=P_{t=0}(-M|E)$, this property stays valid at all times
$P_t(M|E)=P_t(-M|E)$. As stated, this condition of neutrality is quite
stringent. It can be relaxed provided some conditions are met. For
sake of simplicity and to successfully convey the main message of this
paper, we will assume that the observable density $g(M)$ is
symmetric~\footnote{A common choice is a Gaussian distribution of zero
mean and unit variance such that $C(t,t)=1$ holds.},
\be
g(M)=\frac{1}{N}\sum_{\alpha=1}^N \delta(M-M_{\alpha})\equiv g(-M)
\label{a15}
\ee 
Eq.~\eq{a15} together with the initial condition
$P_{t=0}(M|E)=P_{t=0}(-M|E)$ is enough to guarantee neutrality.
Neutrality allows to derive a simple relation for the lowest-time
derivative of the correlation function $C(t,s)$,
\be \frac{\partial C(t,s)}{\partial
s}=\frac{1}{\Delta t}\sum_{\alpha_t,..,\alpha_s}P_s(\alpha_s)M_{\alpha_t}\prod_{r=s}^{t-1}
W(\alpha_{r+1}|\alpha_r) (M_{\alpha_{s+1}}-M_{\alpha_{s}})
\label{a16}
\ee
If we replace \eq{a3} for $W(\alpha_{s+1}|\alpha_s)$ in \eq{a16} then neutrality
implies,
\be
\sum_{\alpha_s}P_s(\alpha_s)A(\alpha_s)M_{\alpha_s}=0
\label{a17}
\ee
because, for a fixed value of the energy, an identical number of traps
contribute to \eq{a17} with positive and negative values of $M$.
Therefore,
\be \frac{\partial C(t,s)}{\partial
s}=\frac{1}{\Delta t}\sum_{\alpha_t,..,\alpha_s}P_s(\alpha_s)M_{\alpha_t}\prod_{r=s+1}^{t-1}
W(\alpha_{r+1}|\alpha_r) A(\alpha_s) M_{\alpha_{s+1}}~~~~.
\label{a18}
\ee
%

\section{The Bouchaud model (BM)}
\label{bm}

In the BM the unperturbed rates $A^{\rm BM}(\alpha'|\alpha)$ are given
by \eq{a6a} and depend only on the energy of the departure trap so we
will adopt the notation $A^{\rm BM}(\alpha'|\alpha)\equiv A^{\rm
BM}(\alpha)$.
Let us consider the rates \eq{a8},
\be
A^{\rm BM}_h(\alpha'|\alpha)=A^{\rm
BM}(\alpha)\exp\Bigl(-\beta h(\mu M_{\alpha}-\gamma M_{\alpha'})\Bigr)\label{b1}
\ee
%
It is easy to verify that in this
case $a(\alpha'|\alpha)$ is given by,
\be
a^{BM}(\alpha'|\alpha)=-\beta A^{BM}(\alpha)(\mu M_{\alpha}-\gamma M_{\alpha'})
\label{b2}
\ee
where $\mu$ and $\gamma$ are two constants which we will assume can take
any value (positive or negative). The parameter $\gamma$ appearing in
the rates \eq{b2} only depends upon the value of the observable $M$ at
the arrival trap. At linear order in $h$, the change of the escape rate
$W^{\rm esc}(\alpha)$ \eq{a5} is given by
\be
W^{\rm esc}_h(\alpha)-W^{\rm esc}(\alpha)\sim - \mu \beta h N A^{\rm
BM}(\alpha) M_{\alpha}+{\cal O}(h^2)=-\omega_0 \mu \beta h M_{\alpha}\exp(\beta
E_{\alpha})+{\cal O}(h^2)
\label{b9}
\ee
and does not depend on $\gamma$. Therefore, while the parameter $\mu$
directly modifies the lifetime of a trap, the parameter $\gamma$ does
not modify it (up to linear order in the perturbation). Simply put,
while the parameter $\mu$ directly affects the height of the energetic
barrier to be surmounted (through thermal activation) to escape
from a trap, the parameter $\gamma$ selectively enhances the rate
toward the destination trap through a different activated mechanism.  
In Sec.~\ref{discussion} we present a physically motivated explanation
for this dependence.

Note that only if $\gamma=1-\mu$ detailed balance is preserved in \eq{b2}. 
Inserting \eq{b2} in \eqq{a12}{a13} we get, 
\be
\Delta_h W(\alpha'|\alpha)=-\beta A^{\rm BM}(\alpha)\Bigl[(\mu M_{\alpha}-\gamma
M_{\alpha'})-N(\mu
M_{\alpha}-\gamma\overline{M})\delta_{\alpha,\alpha'}\Bigr]
\label{b3}
\ee
where $\overline{M}=(1/N)\sum_{\alpha} M_{\alpha}$ is the average
magnetization of traps. 
The following
mathematical steps are followed. We insert \eq{b3} in \eq{a10}
and use the identity \eq{a3}.  If the observable $M_{\alpha}$ is neutral
then from \eq{a15} we have $\overline{M}=0$. Using \eq{a18}
we get,
\be
R(t,s)=-\beta\Bigl[\mu \frac{\partial C(t,s)}{\partial t} -\gamma \frac{\partial C(t,s)}{\partial s} \Bigr]
\label{b4}
\ee
where the continuous limit $\Delta t\to 0$ has been
taken~\footnote{Equation \eq{b4} was first derived in \cite{BouDea95} but for
the case where detailed balance holds. Note also that the symmetry
condition there expressed (our \eq{a15}) is not enough for the
validity of \eq{b4} as full neutrality as expressed by the supplemental condition 
$P_{t=0}(M|E)=P_{t=0}(-M|E)$ must hold.}. Replacing in \eq{a1} we get for
the FDR,
\be X(t,s)=\gamma-\mu\Biggl( \frac{\frac{\partial C(t,s)}{\partial
t}}{\frac{\partial C(t,s)}{\partial s}}\Biggr)
\label{b5}
\ee
Several comments are in order. First, the present analysis covers
regimes where detailed balance is violated. We already justified this
choice (see footnote~\ref{foot1}). In fact, in a time-translational
invariant regime, $X=\gamma+\mu$ which is generally different from
1. Only if detailed balance holds ($\gamma=1-\mu$) then $X=1$. Second,
$X$ is finite in the $T\to 0$ limit. In other words, if a physically
meaningful effective temperature \eq{a1b} exists, then it has to be
proportional to $T$ and vanish at $T=0$.
Let us consider the following aging form of $C(t,s)$,
\be C(t,s)=\int_{0}^{\infty} d\lambda
w(\lambda)\hat{C}_{\lambda}\Bigl(\frac{t-s}{s^{\lambda}}\Bigr)~~~~~;~~~~\int_{0}^{\infty}
d\lambda w(\lambda)=1
\label{b6}
\ee
where $\hat{C}_{\lambda}(0)=1,\hat{C}_{\lambda}(\infty)=0$.
Provided $w(\lambda\to 0)\to 0$ then it can be easily proved that for
$t\gg s$, 
\be X(t,s)\to \gamma-{\cal O}\Bigl( \frac{s}{t}\Bigr)
\label{b7}
\ee
and therefore
\be X_{\infty}=\lim_{t\to\infty} X(t,s)=\gamma
\label{b8}
\ee
i.e. the FDR asymptotically converges to a non-trivial quantity
($\gamma$). The main conclusion of this simple calculation is that for
rates given by \eq{b2} there is a universal dependence of the FDR (in its
long-time limit) on the parameter $\gamma$ describing how transition
rates are perturbed at the linear response level (i.e up to linear order in $h$).
Note that for $\mu=0$,  $X(t,s)=\gamma$ \eq{b5}, the FD plots are full
straight lines as in the 1S scenario and the escape rates are unmodified
at the linear order in the field \eq{b9}.

The validity of \eq{b7} relies on the fact that the system is
aging. Were the system in a stationary state then trivially
$X(t,s)=\gamma+\mu$ for all times $t,s$. Therefore, for stationary
non-equilibrium systems $X_{\infty}=\gamma+\mu$ and an additional
dependence of the limiting value $X_{\infty}$ on $\mu$ is expected.

\begin{figure}
\begin{center}
\epsfig{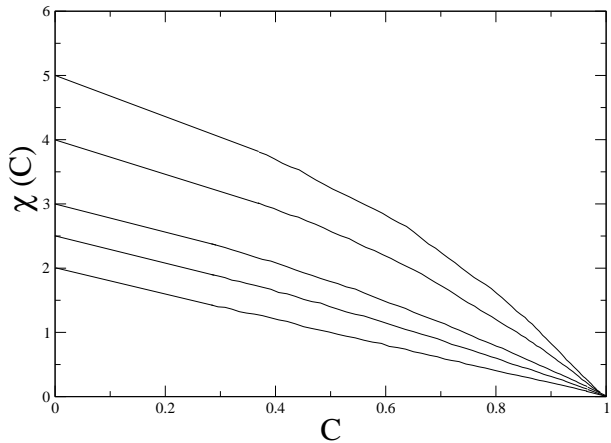}
\end{center}
\caption{FD plots for the BM at $T=0.5$ ($T_g=1$ with $g(M)$ a
Gaussian distribution of unit variance) at $\gamma=1$ and
$\mu=0,0.5,1,2,3$ (from bottom to top). Note that only for $\mu=0$
there is a full straight line, yet in all cases the limiting slope
when $C\to 0$ is the same, $X_{\infty}=\gamma=1$. It can be proved that the value of
$\chi(C)$ in the limit $C\to 0$ is given by $\beta\gamma+\mu/T_g$.}
\label{Fig1}
\end{figure}

In general, for models with full aging $C(t,s)=\hat{C}(t/s)$,  $X(t,s)$
in \eq{b5} is a single function of $C$ yielding limiting FD plots~\cite{BouDea95,FieSol02}. 
For instance, the BM is
characterized by a non-normalizable distribution for $T<T_g$ where the
correlation function $C(t,s)$ shows full aging.
In Fig.~\ref{Fig1} we show FD plots for the BM for
a given value of $\gamma$ and different values of $\mu$. Note that for
$\mu=0$ a  1S FD plot is obtained characteristic of systems
with a single timescale.  The purely activated case $\mu=1,\gamma=0$ gives
$X_{\infty}=0$ as has been found in \cite{FieSol02}. 

\section{The Barrat-Mezard model (BMM)}
\label{bmm}

As we already said in Sec.~\ref{definitions} after \eq{a7b} the main
difference between the BM and the BMM is that the escape rates in the
later are temperature independent in the low $T$ limit, therefore slow
dynamics is governed by entropy barriers. Therefore, the BMM appears as the
phenomenological description of the entropic relaxation of several 
microscopic models such as the Backgammon model~\cite{Ritort95} or the
oscillator model~\cite{BonPadRit98}.

In this section we want to show that similar conclusions \eqq{b9}{b8}
can also be inferred from the BMM \eq{a6b} in the low-temperature limit
$T\to 0$ where a purely entropic mechanism of relaxation is at work (in
contrast to the BM case where escape rates are thermally activated).
The analysis of the FDR in the BMM goes along the same lines as in the
preceding section. We start from the unperturbed $T=0$ rates,
\be
A^{\rm BMM}(\alpha'|\alpha)=\frac{\omega_0}{N}\theta(E_{\alpha}-E_{\alpha'})~~~~.\label{c1}
\ee
We remark two aspects of these rates: 1) As these are zero-temperature
rates they do not satisfy detailed balance and 2) the rates are
independent of the energy of the departure and arrival trap provided
that the energy decreases.  Nevertheless the escape rate \eq{a5} from a
given trap will depend on its energy through the relation,
\be
W^{\rm esc}(\alpha)=\frac{\omega_0}{N}\sum_{\alpha'}\theta(E_{\alpha}-E_{\alpha'})= \omega_0\int_{-\infty}^{E_{\alpha}}\rho(E')dE'
\label{c2}
\ee
and will decrease as the energy of the trap $E_{\alpha}$ decreases.  Now
we need to specify how the rates $W$ change in the presence of a
field. Inspired by the corresponding expression for the BM \eq{b1} we
will assume the following factorized form for the perturbed rates in the BMM
\be A^{\rm BMM}_h(\alpha'|\alpha)=A^{\rm
BMM}(\alpha'|\alpha)\exp\Bigl(-h(\mu M_{\alpha}-\gamma
M_{\alpha'})\Bigr)~~~~.\label{c6} 
\ee
where $A^{\rm BMM}_h(\alpha'|\alpha)$ is given by \eq{c1}. This
functional dependence assumes that transitions to new traps are
enhanced towards larger values of the observable $M_{\alpha'}$,
however arrival traps must have energies below that of the departure
trap $E_{\alpha'}<E_{\alpha}$. This form of the perturbed rates does
not correspond to the case of purely activated rates where \eq{c1} are
modified according to the shift \eq{a7c}. In fact, were this the case
then there would be correlations between energies and observable
values, and transitions with $E_{\alpha'}>E_{\alpha}$ would occur
depending on the difference $M_{\alpha'}-M_{\alpha}$. The factorized
form of the rates \eq{c6} as well as the enhancement of rates towards
traps with larger values of $M_{\alpha}$ are not easy to justify {\em a
priori}. They have to be seen as an approximate solution to the real
perturbed rates that takes into account the dependence of the rates
upon the observable value taken at the arrival trap~\footnote{A
contribution to the perturbed rates of the form \eq{c6} seems to us
essential for the emergence of non-trivial effective
temperatures. After completion of this work, calculations by
Sollich (see note added at the end of the paper) have shown that purely activated rates are
equivalent to \eq{c6} with $\mu=2/T_g,\gamma=0$ where there is no
dependence of the modified rates upon the observable value at the
arrival trap.}.

\begin{figure}
\begin{center}
\epsfig{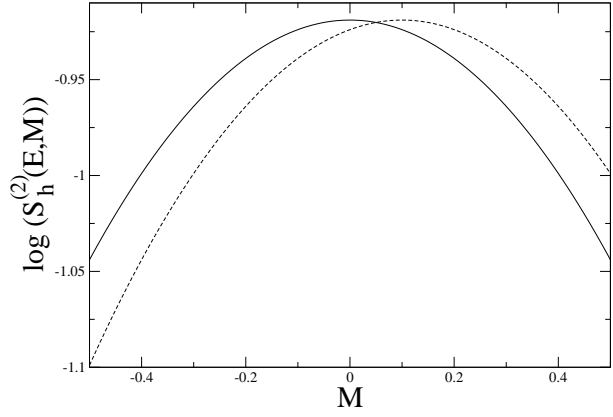}
\end{center}
\caption{Illustrative picture for $S^{(2)}_h(E,M)$ (Gaussian distribution $g(M)$ of unit
variance and $T_g=1$) as a function of $M$ for a fixed value of the energy. $h=0$ (continuous line) and $h=0.1$ (dashed
line). The difference between the two curves is linear with $h$ and the
slope is equal to $h/T_g=0.1$.}  
\label{Fig2}
\end{figure}

\begin{figure}
\begin{center}
\epsfig{file=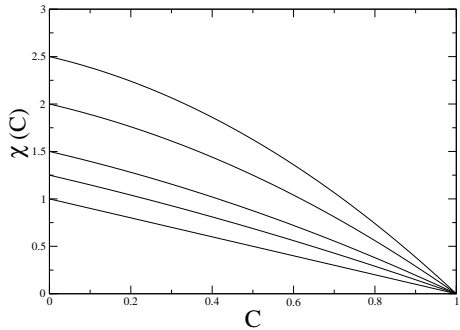,angle=-90, width=7cm}
\end{center}
\caption{FD plots for the BBM at $T=0$ ($T_g=1$ with $g(M)$ a Gaussian
distribution of unit variance) at $\gamma=1$ and $\mu=0,0.5,1,2,3$ (from
bottom to top). Note that only for $\mu=0$ there is a full straight line,
yet in all cases the limiting slope when $C\to 0$ is the same, $T_{\infty}=T_g=\gamma=1$.}
\label{Fig3}
\end{figure}

The enhancement of rates towards traps with larger $M_{\alpha'}$ can
be explained as follows.  Let us assume that $\rho(E)$ is a
monotonically increasing function of $E$ (e.g. as in \eq{a7b}). The
main effect of the field on a shell of traps that have a given energy
is that the population of traps with $M>0$ tends to increase relative
to those with $M<0$. The shift of populations can be computed as
follows. Given the unperturbed density $\rho^{(2)}(E,M)=\rho(E)g(M)$
we can write,
\bea
\rho_h^{(2)}(E,M)=\int_{-\infty}^{0}dE'  \rho(E')g(M)\delta(E-E'-Mh)= \rho(E+Mh)g(M)\\
=\rho(E)g(M)\Bigl(1+\frac{\partial \log(\rho(E))}{\partial E}Mh+{\cal
O}(h^2)\Bigr)=\rho^{(2)}(E,M)\exp\Bigl(\gamma(E) Mh\Bigr) 
\label{c3}
\eea
the final expression being valid up to linear order in $h$ and where we
have introduced the function $\gamma(E)$,
\be
\gamma(E)=\frac{\partial \log(\rho(E))}{\partial E}=\frac{\partial S(E)}{\partial E}~~~~.
\label{c4}
\ee
where $S(E)=\log(\rho(E))$ is an entropy an $\gamma$ has the dimensions
of the inverse of a temperature.  Analogously, if we define an entropy
$S^{(2)}(E,M)=\log(\rho^{(2)}(E,M))$ then we have, up to linear order in
$h$, $S^{(2)}_h(E,M)=S^{(2)}(E,M)+\gamma(E) Mh$.  The schematic behavior
of $S^{(2)}_h(E,M)$ is depicted in Fig.~\ref{Fig2}.  For an
exponential distribution of states \eq{a7b} we have,
\be
\gamma(E)\equiv \gamma=\frac{1}{T_g}
\label{c5}
\ee
Although the density $\rho_h(E,M)$ changes at linear order in $h$, the
symmetry property \eq{a15} ensures that the total density
$\int_{-\infty}^{\infty}\rho_h(E,M)dM$ only changes at the next order
$h^2$.  We stress that the parameter $\gamma$ (expressing the dependence
of the perturbed rates upon the value of the observable evaluated at the
arrival trap), rather than $\mu$, is the only term describing the
entropic unbalance of traps. From \eqq{a5}{c6} we obtain for the escape
rates, at linear order in the field,
\be
W^{\rm esc}_h(\alpha)=\omega_0\exp\Bigr( \frac{E_{\alpha}-h \mu M_{\alpha}}{T_g}\Bigl)
\label{c6b}
\ee
where we used the result $\overline{M}=0$ and \eq{c2}. Calculations identical to
those shown for the BM in Sec.~\ref{bm} lead to,
\be
R(t,s)=-\Bigl[\mu \frac{\partial C(t,s)}{\partial t} -\gamma \frac{\partial C(t,s)}{\partial s} \Bigr]
\label{c7}
\ee
Note that, contrarily to \eq{b4}, the response is now finite in the $T=0$
limit, therefore the FDR vanishes linearly with $T$ as $T\to 0$.  Rather
than $X$ it is useful to consider the effective temperature \eq{a1b},
\be
T_{\rm eff}(t,s)=\Biggl[\gamma-\mu\Biggl( \frac{\frac{\partial C(t,s)}{\partial
t}}{\frac{\partial C(t,s)}{\partial s}}\Biggr)\Biggr]^{-1}
\label{c8}
\ee
Again, under the same assumptions \eq{b6} and using \eq{c5} we get in the limit $t\to \infty$,
\be T_{\infty}(s)=\lim_{t\to\infty} T_{\rm eff}(t,s)=\frac{1}{\gamma}=T_g
\label{c9}
\ee
which is independent of $s$ and $\mu$. Therefore, in the limit $t\to \infty$ 
the effective temperature coincides with $T_g$ which is the width of
the exponential distribution of states through \eqq{c4}{c5}.
The change of the escape rate \eq{a5} induced by the field is given by
\be W^{\rm esc}_h(\alpha)-W^{\rm esc}(\alpha)=-\omega_0 \mu h
M_{\alpha}\exp\bigl(\frac{E_{\alpha}}{T_g}\bigr)+{\cal O}(h^2)
\label{c10}
\ee
and the escape rate, as for the BM, only depends on the value of
$\mu$. Again, full straight FD plots are obtained for $\mu=0$ when the
escape rates (or trapping times) are unchanged by the field.  For the
BMM, computations of \eqq{c8}{c9} can be made explicit due to the
simplicity of the expression for the correlation function, $C(t,s)=s/t$
for $t,s\gg 1$~\cite{BMM}.  For the integrated response function we get,
\be
\chi(t,s)\equiv \chi(C)=\int_s^t\,dt'R(t,t')=\gamma (1-C)+\frac{\mu}{2}(1-C^2)
\label{c12}
\ee
and the effective temperature,
\be
\frac{1}{T_{\rm eff}(C)}=-\frac{\partial \chi(C)}{\partial C}=\gamma+\mu C
\label{c11}
\ee
giving for $\mu=0$ straight FD plots characteristic of 1S behavior,
$\chi(C)=\gamma (1-C)$. Note that, in the $\mu=0$ case,
$\chi(C=0)=\gamma=1/T_g$ which is the value of the 
susceptibility at the transition temperature. Fig.~\ref{Fig3} shows some typical FD
plots.

\subsection{Numerical test of the BMM solution with rates \eq{c6}.}
\begin{figure}
\begin{center}
\epsfig{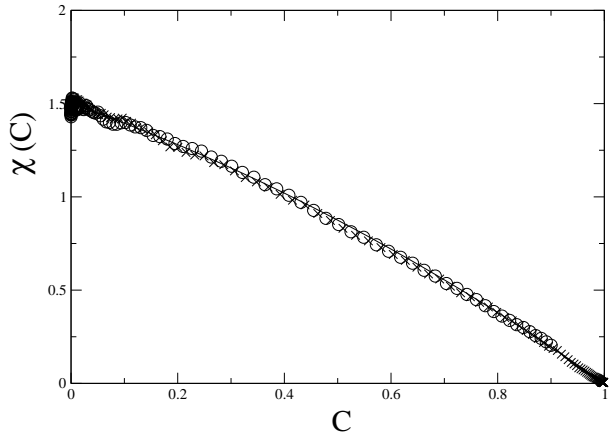}
\end{center}
\caption{Numerical test of \eq{c12}. FD plots for the BBM with
$\gamma=\mu=T_g=1$, $\omega_0=1$ and two waiting times $s=10$
(circles) and $s=1000$ (crosses) in the asymptotic regime
$s>>1/\omega_0$.  The continuous line is the analytical result
\eq{c12}.}
\label{Fig4}
\end{figure}

To verify the correctness of the previous solution we have done a
numerical simulation of the BMM with rates \eq{c6}. We have considered a
Gaussian distribution of observables $g(M)=(2\pi)^{-1/2}\exp(-M^2/2)$
and $T_g=1$. Let $E,M$ be the energy and trap at the departure state. The procedure
is as follows: 1) a trap with observable $M'$ is chosen with a
probability proportional to $\exp(M'h/T_g)g(M')$, i.e. $M'=r +h/T_g$ where
$r$ is a Gaussian (zero mean and unit variance) random number; 2) the
lifetime of the trap, according to \eq{c6b}, is given by
$\tau=(1/\omega_0)\exp(-E+Mh)$; 3) The arrival energy is given by
$E'=E+\log(r')$ where $r'$ is a random number uniformly distributed
between 0 and 1. We start from the disordered initial state $E=0$, $M=r$
where $r$ is a Gaussian random number as before. We then run two
independent simulations by iterating many times the previous steps (one
run is at $h=0$ for all times, the other run is at $h=0$ until time $s$
and later a small field $h$ is switched on) and generate correlations
and susceptibilities \eq{c12} at different times $t,s$. For
$t,s>1/\omega_0$ the asymptotic aging solution is obtained as well as
the limiting FD plots \eq{c12}. We verified the correctness of the
solution $C(t,s)$ at zero field. The numerical results are shown in
Fig.~\ref{Fig4} (right) together with the analytical prediction \eq{c5}
with $\gamma=\mu=1$ (continuous line). 
\section{Discussion}
\label{discussion}
The entropic mechanism described in the preceding section for the BMM
provides an explanation on the emergence of an effective
temperature given by the width of the exponential density of trap
energies~\footnote{In the mean-field theory of spin glasses such origin
of the effective temperature has been already
remarked~\cite{MezParVir85,FraVir00}. Yet in these models activated
processes are completely suppressed.}. However, is it possible to extend that explanation to
the BM where such entropic mechanism appears difficult to reconcile with
the explicit activated behavior of the rates \eq{b1}?  
In fact, there is an important difference between the BM and the
BMM. While in the later the perturbed rates are not thermally activated
(see \eq{c6}) leading to a value of $X_{\infty}$ proportional to $T$ at
low $T$, in the former the perturbed rates are thermally activated
\eq{b1} leading to a value of $X_{\infty}$ that is temperature
independent at low $T$. 

The following explanation, albeit speculative, might be useful. Consider
a glass where relaxation is driven by the nucleation of cooperatively
rearranging regions. The nucleation of such a region requires the
simultaneous activation of $M$ independent processes, each one
characterized by a thermally activated rate $w$ \footnote{Several
scenarios based on this cooperative behavior have been
proposed~\cite{CriRit01,ChaBag03} to explain the origin of
super-activated effects in structural glasses.}.  In the presence of a
perturbation of intensity $h$ the rates get modified $w_h\sim
w\exp(-\beta h \mu)$, where $w$ is the rate at zero field and $\mu$ is a
microscopic parameter. The modified rate for the nucleation of the
region is $w_h^M\sim w^M \exp(-\beta h \mu M)$. If $\alpha,\alpha'$ denote
the configurations of the system before nucleating and after nucleating
respectively, then $M$ can be identified with the net change of the
observable $M_{\alpha}-M_{\alpha'}$ after nucleation takes place. The
overall rate for that nucleating process is $\exp(-\beta h \mu
(M_{\alpha'}-M_{\alpha}))$ which coincides with \eq{b1} with
$\gamma=\mu$. 

This scenario could describe diffusive processes (such as low $T$
coarsening) where partial equilibration occurs in the short-time sector
corresponding to relaxation within the bulk of domains. In this case, as
$X(t,s\to t^-)\to \gamma+\mu\to 1$, the identity $\gamma=\mu$
gives $X_{\infty}=\gamma=\mu=1/2$, a result that has been found in the
one-dimensional Glauber-Ising
chain~\cite{GodLuc00,LipZan00,FieSol02b}. Let us mention that FD plots
similar to those shown here (Figs.~\ref{Fig1}, \ref{Fig3}) with a
limiting value $X_{\infty}=1/2$ have been reported in one-dimensional
models~\cite{FieSol02b}, yet it would be interesting to rigorously
explain their emergence along the present line of reasoning.
The conclusion we draw from this paragraph is that a scenario where
$X_{\infty}$ is finite at $T=0$ does not appear compatible with an
entropically driven relaxation.

The validity of the expression $w_h\sim w\exp(-\beta h \mu )$ requires
that all rates corresponding to all elementary activated processes are
modified by the same amount (up to linear order in $h$) by the
perturbation, a requirement that may not be always fulfilled. For
instance, let us consider a spin-glass Ising system staying in a locally
stable energy minima while aging (i.e. the configuration is stable
against all one-spin flips). Assume also that $P$ spins (pointing in
different directions) need to be simultaneously activated for the system
to escape from that local minima. If there is no long-range order (as is
common in spin glasses) the local field distribution is symmetric around
zero, so the local fields acting on these $P$ spins will point in
different directions. A uniform perturbing field will therefore increase
or decrease the local field, and the modified rates $w_h\sim
w\exp(-\beta h\mu)$ will not be the same for all $P$ spins. Therefore,
the mechanism behind the emergence of effective temperatures in {\em
frustrated} glassy systems (as compared to coarsening systems) seems
better described by the BMM rather than by the BM.

There are two additional aspects we would like to discuss.  One aspect
concerns the predicted waiting-time dependence of $X_{\infty}$ and
$T_{\rm eff}$ in 1S models. For trap models described by a
time-independent density of traps \eq{a7b} such dependence is not
expected, indeed the limiting values \eqq{b8}{c9} are time
independent. For generic glassy systems, though, the width of the
exponential density of traps (i.e. $\gamma$ according to \eq{c5}) should
be time dependent as the average energy decreases. Including such
dependence would produce FD plots identical to those found in other
solvable models like the oscillator model~\cite{BonPadRit98}.  The
second aspect concerns the characteristic form of FD plots in realistic
glass models.  How to reproduce 1S FD plots with two linear segments,
one for $C\to 1$, the other for $C\to 0$? This requires the introduction
of some internal dynamics within traps which is absent in the original
formulation of both the BM and the BMM. In the same vein, generalization
of this treatment beyond the 1S pattern to many timescales could be
accomplished through more general transition rates depending on trap
correlations $q_{\alpha,\alpha'}=M_{\alpha}M_{\alpha'}$ in some type of
hierarchical structure~~\cite{BouDea95}.

Finally some words about the violation of the property of detailed
balance that we have tacitly assumed in the rates \eqq{b1}{c6}. Detailed
balance is of course violated in the BMM at $T=0$. As we have already
insisted on (footnote~\ref{foot1}) this assumption is essential, as trap
models are coarse-grained descriptions in phase space of a microscopic
dynamics that satisfies that property. The rates $W(\alpha'|\alpha)$ in
\eq{a2} are obtained by projection of the microscopic probabilities and
rates onto the corresponding coarse-grained representation of the
original phase space. In this projected description detailed balance is
not fulfilled anymore (for a discussion see Sec.3.3 in \cite{CriRit03}).
Moreover, detailed balance is known to be violated in driven systems,
yet they also show 1S FD violating patterns. Therefore, an explanation
on the emergence of effective temperatures cannot rely on the
fulfillment of the detailed balance property but need to be more
general. A coarse-grained description of glassy dynamics, where rates
are entropically driven, accomplishes this goal.

\section{Concluding remarks}
\label{conclusions}

Trap models represent a useful approach to understand universal
aspects of glassy dynamics. They can be seen as a coarse-grained
description of structural relaxation where nucleation of cooperatively
rearranging regions is modeled by jumps among traps. They are
expected to be relevant to describe relaxation processes in structural
glasses. The vast majority of theoretical and numerical studies on
glassy systems show that one-step (1S) FD plots are common in the
aging regime. However, recent studies on trap models have not found
indications of 1S behavior in FD plots, casting doubts on the
suitability of trap models to reproduce this particular feature of
glassy systems.

In this paper we have addressed this issue by showing that proper
treatment of the influence of the perturbation on the transition rates
partially resolves the problem. Indeed, 1S plots can be obtained if entropic
effects, driven by the unbalance of the population of traps in the
presence of a field, are included in the analysis. We concentrated our
attention in both the Bouchaud model (BM) and the Barrat-Mezard model
(BMM), in this last case at zero-temperature to better emphasize the
main result. We find that $X_{\infty}=\lim_{t\to \infty}X(t,s)$ (for the
BM) and $T_{\infty}=\lim_{t\to \infty}T_{\rm eff}(t,s)$ (for the BMM at
$T=0$) are universal quantities that only depend on a parameter
($\gamma$) describing the dependence of the perturbed rates on the
observable value at the arrival trap. Yet, the physical origin of this
dependence appears very different in both models. While perturbed rates
are activated in the BM, they are not in the BMM but originate from the
entropic unbalance of traps upon the action of the field. This shows
that 1S behavior in glassy systems, and the emergence of finite
effective temperatures in the low $T$ limit, are the fingerprint of a
dynamics governed by entropic relaxational processes. Straight 1S plots
(compatible with a unique long-time sector) are obtained in both models
only when escape rates are unchanged (up to the linear order) by the
action of the field.

Which model (BM or BMM) represents more faithfully the
slow dynamics of 1S systems?  The most probable answer is that neither
the BM or the BMM provides a complete qualitative description of glassy
relaxation and that aspects of both models must simultaneously
intertwine in the overall relaxation. The dependence of the perturbed
rates on the observable taken at the departure trap (the term including
the parameter $\mu$ in \eq{b1}) can be identified with energy barriers
governing the thermally activated escape process from a trap. While the
dependence on the observable taken at the arrival trap (the term
including the parameter $\gamma$ in \eq{c6}) is consequence of the
unbalance of the number of traps induced by the field, leading to
temperature independent entropic relaxation.  Also, the full
straight-line FD plots obtained for $\mu=0$ represent an extremely
simplified behavior of a more realistic scenario where a dependence of
the trapping time at linear order in the field is expected. Yet, this
dependence is not in conflict with the emergence of ``quasi-straight''
FD plots for which the existence of a limiting value
$T_{\infty}(s)$ is enough.

There remain several open question for the future. It could be
instructive to extend this analysis to the BMM at finite
temperature.  However, a relevant question is to
understand how all these considerations are to be generalized to include
the case of arbitrary observables. For the non-neutral case, explicit
computations as have been done here are not possible. Only for the
purely activated case calculations can be done~\cite{FieSol02}, but that
analysis does not include the entropic effects discussed in this
paper. Yet, there may exist general proofs from which one can deduce
specific predictions to be challenged in several microscopic theoretical
models and well as in numerical simulations.

{\bf Acknowledgments.} I warmly thank P. Sollich for communicating me
his results after submission of the first draft of the paper and the
fruitful exchange of ideas that followed after. FR is supported by the
Spanish Ministerio de Ciencia y Tecnolog\'{\i}a Grant BFM2001-3525 and
Generalitat de Catalunya. FR has benefited from the warming
hospitality at the Bustamante lab in the University of California in
Berkeley where this work has been done.

{\em Note added: Just after completion of this worked I learned of
calculations by Sollich showing that, for the BMM,
purely activated rates \eq{a7c} lead to FD plots not displaying 1S
behavior, yet they seem to be described by \eq{c6} with
$\gamma=0,\mu=2/T_g$, i.e. without dependence of the rates upon the value
of the observable taken at the arrival trap. FD plots
are therefore very similar to those found in the BM with purely activated
rates. This result is very interesting as it shows that the
emergence of 1S behavior in trap models cannot be described by only
shifting the energies according to the value of the
observable [cfr. \eq{a7c}]. Identification of mechanisms behind the
emergence of perturbed rates such as \eqq{b1}{c6}, beyond the purely
activated ones, is certainly needed to clarify this issue.}


\begin{thebibliography}{99}



\bibitem{BM} J.-P. Bouchaud, J. Physique I, {\bf 2}, 1705 (1992)

\bibitem{BMM} A. Barrat and  M. Mezard, J. Physique I, {\bf 5}, 941 (1995)

\bibitem{BouCugKurMez98} J. P. Bouchaud, L. F. Cugliandolo, J. Kurchan and M. Mezard in 
{\em Spin Glasses and Random Fields} A. P. Young (Ed), World
Scientific, Singapore (1998)

\bibitem{BouDea95} J.-P. Bouchaud and D. S. Dean, J. Physique I, {\bf 5}, 265 (1995)

\bibitem{BouComMon95} J.-P. Bouchaud, A. Comtet and C. Monthus, J. Physique I, {\bf 5}, 1521 (1995); 
C. Monthus and J.-P. Bouchaud, J. Phys. A (Math. Gen.) {\bf 29}, 3847 (1996)


\bibitem{CriRit03} A. Crisanti and F. Ritort, 
Preprint {\bf condmat}/0212490 to appear in J. Phys. A

\bibitem{Cugliandolo02} L. F. Cugliandolo, Preprint {\bf condmat}/0210312

\bibitem{GodLuc00a} C. Godreche and J. M. Luck, J. Phys. A (Math. Gen.) {\bf 33}, 1151 (2000)

\bibitem{LipZan00} E. Lippiello and M. Zannetti, Phys. Rev. E {\bf 61}, 3369 (2000)


\bibitem{CalGam02} P.Calabrese and A. Gambassi, Phys. Rev. E {\bf 65}, 066120 (2002)

\bibitem{FieSol02} S. M. Fielding and P. Sollich, Phys. Rev. Lett. {\bf 88}, 050603 (2002)

\bibitem{FieSol02b} P. Sollich{,} S. Fielding and P. Mayer, J. Phys. C (Cond. Matt.), {\bf 14}, 1683 (2002).


\bibitem{BerBou02} E. Bertin and J.-P. Bouchaud, J. Phys. A (Math. Gen.) {\bf 35}, 3039 (2002)


\bibitem{Ritort95} F. Ritort, Phys. Rev. Lett., {\bf 75}, 1190 (1995)


\bibitem{BonPadRit98} L. L. Bonilla, F. G. Padilla and F. Ritort, Physica A, {\bf 250}, 315 (1998)


\bibitem{MezParVir85} M. Mezard, G. Parisi and M. A. Virasoro,
J. Physique. Lett. {\bf 46}, L217 (1985) 

\bibitem{FraVir00} S. Franz and M. A. Virasoro, J. Phys. A (Math. Gen.)
{\bf 33}, 891 (2000)

\bibitem{CriRit01} A. Crisanti and F. Ritort, Preprint {\bf condmat/0102104}

\bibitem{ChaBag03} D. Chakrabarti and B. Bagchi, Preprint  {\bf condmat/0303153}

\bibitem{GodLuc00} C. Godreche and J. M. Luck, J. Phys. A (Math. Gen.) {\bf 33}, 1151 (2000)





\end{thebibliography}
\end{document}